# Infrared Spectroscopy of Cyanonaphthalenes under Interstellar Relevant Conditions and Their Potential Connection with Astronomical Aromatic Infrared Bands

Jiaqi Xin[1], Jianzhi Xu[1], Piero Ferrari[2], Gao-Lei Hou[1,*]

[1]MOE Key Laboratory for Non-Equilibrium Synthesis and Modulation of Condensed Matter, School of Physics, Xi´an Jiaotong University, Xi´an, 710049 Shaanxi, China

[2]HFML-FELIX, Nijmegen, The Netherlands

E-mail: gaolei.hou@xjtu.edu.cn

## ABSTRACT

*Context.* Aromatic infrared bands (AIBs) are widely observed in diverse astrophysical environments and are generally attributed to vibrational emission from polycyclic aromatic hydrocarbons (PAHs). The recent interstellar detection of 1-cyanonaphthalene (1-CNN) and 2-cyanonaphthalene (2-CNN) has motivated detailed infrared spectroscopic studies of cyano-substituted PAHs.

*Aims.* We aim to characterize the infrared spectra and vibrational modes of neutral 1-CNN and 2-CNN under cold and gas-phase conditions and to assess their possible spectroscopic relevance to the astronomical AIBs.

*Methods.* The gas-phase infrared spectra of neutral 1-CNN and 2-CNN were measured in a cold molecular beam using ion-dip spectroscopy. The observed bands were assigned with the aid of harmonic and anharmonic calculations at the B3LYP/N07D level. Infrared emission spectra were subsequently simulated from the experimental spectra within a single-photon approximation framework.

*Results.* We report the infrared spectra of neutral 1-CNN and 2-CNN measured under cold and gas-phase conditions relevant to the interstellar medium. Their vibrational features were assigned in detail, including fundamental vibrations as well as overtone and combination bands. The simulated emission spectra exhibit features in several wavelength regions associated with prominent AIBs, including the aromatic CH stretching region near 3.3 µm, the CC stretching region near 6.2 µm, the mixed CH in-plane bending and CC stretching region at 8.6–8.9 µm, and the CH out-of-plane bending region between 10 and 15 µm.

*Conclusions.* The present spectra provide laboratory reference data for small cyano-substituted PAHs and offer useful clues for interpreting selected AIB regions. These results suggest that cyanonaphthalene molecules are promising contributors to the aromatic infrared bands.

## 1. Introduction

Since the 1970s, a set of prominent infrared emission features have been ubiquitously observed at wavelengths of approximately 3.3, 6.2, 7.7, 8.6, 11.3, and 12.7 μm in a wide range of astrophysical environments, including planetary nebulae, reflection nebulae, and the diffuse interstellar medium of the Milky Way (Gillett et al. 1973; Russell et al. 1977; Bregman et al. 1989; Kwok et al. 1999; Uchida et al. 2000; Peeters 2011). These features are found to be accompanied by numerous weaker emission bands, which together account for approximately 20% of the total infrared emission of the Galaxy (Smith et al. 2007; Tielens 2008). Because the carriers of these features were not identified at the time of their discovery, they were initially called unidentified infrared emission (UIE) bands (Allamandola et al. 1985; Tielens 2008). Subsequent progress in laboratory spectroscopy and theoretical calculations suggested that these major features are consistent with the characteristic vibrational modes of aromatic hydrocarbons (Szczepanski & Vala 1993; Cook et al. 1996; Langhoff 1996; Allamandola et al. 1999; Bauschlicher Jr & Bakes 2000; Bauschlicher Jr et al. 2008). In particular, the UIE bands at 3.3, 8.6, 11.3, and 12.7 μm are suggested to correspond to the vibrational modes of aromatic CH bonds, and the 6.2 and 7.7 μm features are attributed to the vibrational modes of aromatic CC bonds (Knacke 1977; Duley & Williams 1981; Leger & Puget 1984; Allamandola et al. 1985). Thus, these infrared features were later called aromatic infrared bands (AIBs). Identifying the exact carriers of the AIBs and understanding their formation and evolution in different astrophysical environments are of fundamental importance for elucidating the carbon cycle in space, the history of star formation, galaxy evolution, and processes relevant to the origin of life (Tielens 2013; Kwok 2017; Li 2020).

A variety of candidate carriers have been proposed to explain the UIE bands (Sakata et al. 1984; Borghesi et al. 1987; Duley et al. 1989; Papoular et al. 1989; Balm & Kroto 1990; Jones & d'Hendecourt 2000; Kwok & Zhang 2011; Yang et al. 2017a; Hou et al. 2023), among which the polycyclic aromatic hydrocarbon (PAH) hypothesis was the most widely accepted interpretation, attributing the AIBs primarily to the vibrational transitions of PAH molecules excited by single-photon absorption (Leger & Puget 1984; Allamandola et al. 1985; Puget & Léger 1989; Schlemmer et al. 1994; Peeters et al. 2002). This mechanism, proposed by Andriesse (1978) and Sellgren (1984), suggests that PAHs undergo transient heating upon absorbing a single ultraviolet photon and subsequently relax through infrared fluorescence, producing the observed infrared emissions. In recent years, laboratory experiments and theoretical calculations provide further support for the PAH hypothesis (Szczepanski & Vala 1993; Cook et al. 1996; Langhoff 1996; Allamandola et al. 1999; Bauschlicher Jr & Bakes 2000; Bauschlicher Jr et al. 2008; Mattioda et al. 2020; Banhatti et al. 2021; Mackie et al. 2021; Rap et al. 2022). For example, Tielens and co-workers employed low-temperature matrix-isolation Fourier Transform Infrared (FTIR) spectroscopy and PAH emission modelling to show that ionized PAHs strongly enhance the 6–9 μm vibrational bands, providing key experimental evidence that PAH cations contribute significantly to the astronomical AIB features (Allamandola et al. 1989b; Salama & Allamandola 1992; Hudgins et al. 1994; Hudgins & Allamandola 1999b). Li and co-workers performed a series of investigations into the chemical structures and emission spectra of potential AIB carriers, with particular emphasis on the relative strengths of aromatic and aliphatic components (Draine & Li 2007; Yang et al. 2017a; Yang et al. 2017b). Most recently, Brünken and co-workers studied a wide range of small PAH molecules employing infrared multiple photon dissociation (IRMPD) spectroscopy and explored their potential connections to the AIBs (Banhatti et al. 2021; Rap et al. 2022; Rap et al. 2023; Rap et al. 2024a; Rap et al. 2024b). Besides, Esposito et al. (2025) investigated the CN stretching vibration of gas-phase neutral 9-cyanoanthracene through infrared absorption spectra combined with anharmonic quantum chemical calculations, highlighting the diagnostic value of the CN stretch for identifying cyano-PAHs in astronomical environments.

Despite the widespread support for the PAH hypothesis, unambiguous astronomical identification of specific PAH molecules remains a great challenge since 1980s (Tielens 2008; Yang et al. 2017a). In 2021, McGuire et al. (2021) conducted radio observations of the cold Taurus Molecular Cloud-1 using the Green Bank Telescope and reported the detection of rotational transitions of 1-cyanonaphthalene (1-CNN) and 2-cyanonaphthalene (2-CNN), marking the first identification of a specific PAH molecule in space. Detection of these cyano-substituted PAHs at radio frequencies suggests that they may also be observable at other wavelengths, for example in the infrared region. Thus, an interesting question arises, if cyanonaphthalenes could be part of the carriers of the long-thought AIBs? To our knowledge, there exist several theoretical and experimental studies on the infrared spectral properties of CNNs in the literature (Bauschlicher Jr 1998; Esposito et al. 2024; Rasmussen et al. 2025). Ramachandran et al. (2023) reported infrared spectra of 1-CNN and 2-CNN ices in the spectral region of 600–3500 $cm^{-1}$ and assigned several mid-infrared bands with the aid of theoretical calculations. Rawat et al. (2025) investigated the vibrational spectra of 1-CNN and 2-CNN in solid states using infrared and Raman spectroscopy. These studies have greatly advanced our knowledge on the vibrational structures of CNNs in condensed phase. More recently, Bentley et al. (2026) reported the room-temperature gas-phase far-infrared and rotational spectra of 1-CNN and 2-CNN, providing spectroscopic information in the low-frequency region. However, gas-phase experimental study on the infrared spectra of neutral CNNs in the mid-infrared region, which are more relevant in order to compare with the AIBs, have been reported.

In this work, gas-phase infrared spectra of 1-CNN and 2-CNN were measured in a cold molecular beam that is relevant to the interstellar conditions, by means of ion-dip spectroscopy, and assigned with the help of theoretical calculations. We also simulate the infrared emission spectra of the studied CNNs using the experimental infrared spectra to directly compare with astronomical AIBs. Detailed analysis of the correspondence between the CNN emission spectra and the AIBs was performed, providing support for the potential contributions of CNN molecules to AIBs.

## 2. Methods

### *2.1. Experimental methods*

A molecular beam of neutral 1-CNN (2-CNN) molecules seeded in high purity Ar is formed by laser desorption, in which a roughly 1:1 mixture of 1-CNN (2-CNN) and black carbon powder are pressed on the surface of a graphite bar that is placed in front of a pulsed gas valve. The 1064 nm light of a Nd:YAG laser is used to desorb the 1-CNN (2-CNN) molecules from the graphite bar, which are merged with an expanding Ar gas injected by the pulsed gas valve. After formation, the molecular beam of neutrals is ionized using the 193 nm photons from an ArF laser, in a (1+1) resonant enhanced multiple photon ionization (REMPI) scheme. Under these cold gas-phase conditions, the exact temperature of the molecular ensemble cannot be directly determined; however, based on previous studies performed under comparable experimental conditions, the molecular temperature is estimated to be in the range of 20–40 K (Lemmens et al. 2023). After ionization, the composition of the molecular beam is characterized using a reflectron time-of-flight mass spectrometer. More details of the experimental apparatus can be found elsewhere (Ferrari et al. 2024). The infrared spectrum is measured using ion-dip spectroscopy (Bakels et al. 2020), in which the infrared light from the free electron laser FELIX (Nijmegen, The Netherlands) is coupled with the molecular beam and upon excitation with a specific vibrational mode, induces a decrease in the ionization efficiency of the molecule, thus resulting in a signal decrease measured in the mass spectra. For the current experiments, FELIX is tuned in the ranges of 350–1000 $cm^{-1}$ and 780–2200 $cm^{-1}$. The infrared light from FELIX is merged with the molecular beam in a counter-propagating way and is timed to interact with the beam 300 μs prior to ionization. During the experiments, FELIX is scanned in steps of 2.5 $cm^{-1}$. Due to the suboptimal overlap between the two spectral segments of 1-CNN in the 900–1000 $cm^{-1}$ region, the data acquired over the 780–2200 $cm^{-1}$ range were used for this interval to ensure reliable peak identification. To account for the variations in laser power along the scanned wavelength range, the infrared intensity is defined as $I_{IR} = -\ln(I/I_0)/P$, where $I$ and $I_0$ are the intensity of the 1-CNN (2-CNN) peak in the mass spectra with and without the infrared light, respectively, and $P$ is the wavelength-dependent laser power.

*2.2. Computational methods*

Geometry optimization and infrared spectra calculations for 1-CNN and 2-CNN were performed using the Gaussian 16 (Frisch 2019) program. The molecular geometries were optimized using density functional theory (DFT) at the B3LYP/N07D level of theory. Harmonic frequency calculations were performed to verify that all the optimized geometries are minima on the potential energy surfaces and to assign fundamental vibrational modes. The calculated harmonic frequencies were corrected using a scaling factor of 0.976 (Rap et al. 2022) for comparison with the experimental spectra. To further interpret possible non-fundamental features in the experimental spectra and to improve the simulation of vibrational spectra in terms of both band positions and relative intensities, anharmonic calculations were carried out using second-order vibrational perturbation theory (VPT2) (Bloino et al. 2016; Xin et al. 2025). Since the high frequency region often involves overtone and combination bands, as well as anharmonic effects associated with fundamental modes, the experimental bands above 1650 $cm^{-1}$ were assigned on the basis of the anharmonic calculations. These calculations were also carried out at the B3LYP/N07D level, which has been shown to provide reliable anharmonic frequency predictions (Chen 2018; Peeters et al. 2021). All calculated stick spectra were convolved with a Lorentzian function using a full width at half maximum of 8 $cm^{-1}$ (Lu & Chen 2012; Lu 2024).

## 3. Results

*3.1. 1-CNN infrared spectrum*

The experimental and theoretical infrared spectra of 1-CNN are presented in Fig. 1(a), with the assignments of selected fundamental vibrational modes of relatively strong intensity listed in Table 1. More comprehensive assignments of fundamental vibrational modes are provided in Table A.1. All assigned vibrational modes are explicitly labeled in Fig. 2. Because the experimental spectra in the 2000–2200 cm$^{-1}$ region exhibit relatively high noise levels and poor signal quality, this region is excluded from the following discussion for both 1-CNN and 2-CNN. In the 350–1650 cm$^{-1}$ region, the calculated harmonic spectrum exhibits good agreement with the experiment, with the calculated harmonic frequencies scaled by a factor of 0.976. Based on the origin of the vibrational modes, this spectral range can be divided into three sections: in the 350–700 cm$^{-1}$ region, the predicted bands are attributed to CC in-plane and out-of-plane bending modes; in the 700–1000 cm$^{-1}$ region, the predicted bands are dominated by CH out-of-plane bending modes; while in the 1000–1650 cm$^{-1}$ region, CH in-plane bending and CC stretching vibrations are predominant. Several strong characteristic peaks are observed at 447, 769, 798, and 1515 cm$^{-1}$. Specifically, the 447 cm$^{-1}$ peak is assigned to a CC out-of-plane bending mode, the 769 and 798 cm$^{-1}$ peaks both originate from the combined contributions of CC and CH out-of-plane bending modes, and the 1515 cm$^{-1}$ peak is collectively caused by CH/CC in-plane bending and partial CC stretching vibrations.

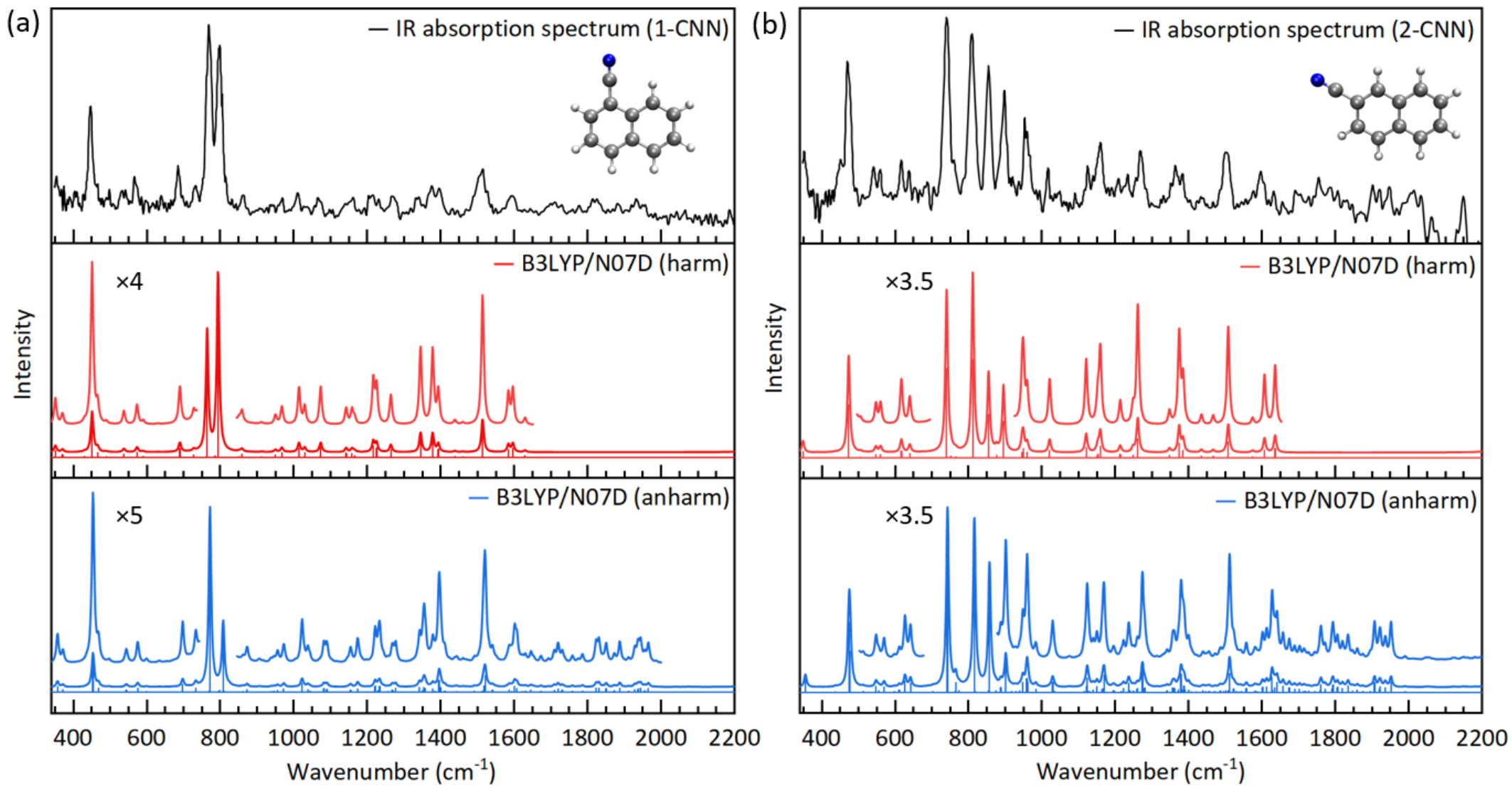

**Fig. 1.** Experimental and calculated infrared spectra as well as the chemical structures of CNN isomers: (a) 1-CNN, (b) 2-CNN. The black curves represent the experimental gas-phase infrared spectra of CNNs, the red curves show the calculated harmonic infrared spectra, and the blue curves correspond to the calculated anharmonic infrared spectra.

Not all calculated fundamental frequencies manifest as distinct features in the measured infrared spectrum. For example, a theoretical feature predicted at 787 cm$^{-1}$ ($\nu_{18}$) is visible in the stick representation of the calculated spectrum (Figure 1), but due to the limited resolution and the broadening applied to the theoretical curve, it is not resolved in either the broadened theoretical spectrum or the experimental measurement. The calculated features at 1159 cm$^{-1}$ ($\nu_{30}$) and 1166 cm$^{-1}$ ($\nu_{31}$) are represented by a single experimental peak at 1162 cm$^{-1}$. The predicted 1585 cm$^{-1}$ ($\nu_{41}$) and 1597 cm$^{-1}$ ($\nu_{42}$) modes, which exhibit similar frequencies and intensities, are likely merge into a single peak at 1596 cm$^{-1}$ in the experimental spectrum. Detailed information for all the vibrational modes mentioned above is provided in Table A.1.

In the 1600–2000 cm$^{-1}$, anharmonic calculations were employed to provide a more reliable interpretation of the experimental spectrum, showing good agreement within this range. Combination and overtone transitions are forbidden under the harmonic approximation; the inclusion of anharmonic correction is essential for their accurate calculations. In the present study, we assigned the experimental features in this region (details are available in Table A.2 of Appendix A). The black arrows in Fig. 2 indicate the bands assigned as overtone and combination transitions. The agreement between the experimental spectrum and anharmonic calculations confirms that these features represent genuine vibrational transitions. Compared with previous assignments based on FTIR spectroscopy (Rawat et al. 2025), several vibrational modes have been revised: the peaks at 769 cm$^{-1}$ ($\nu_{17}$) and 798 cm$^{-1}$ ($\nu_{19}$) are both assigned to CH and CC out-of-plane bending, which differs from earlier FTIR studies that assigned them to CC in-plane bending and CH in-plane bending, respectively. This assignment is supported by the previous investigation of 1-CNN in low-temperature ice phases. Detailed information regarding various vibrational modes is provided in the Appendix A. A comparison with the recently reported gas-phase FTIR spectrum of 1-CNN (Bentley et al. 2026) shows overall good agreement for the commonly observed low-frequency modes. The present bands at 355, 447, 466, 542, and 567 cm$^{-1}$ correspond closely to the reported features at 353, 447, 462, 539, and 571 cm$^{-1}$, respectively, with deviations generally within 4 cm$^{-1}$. In the region near the calculated harmonic mode at 370 cm$^{-1}$, two experimental bands are observed at 368 and 375 cm$^{-1}$. Although the 368 cm$^{-1}$ band is closer to the gas-phase FTIR feature reported by Bentley et al. (366 cm$^{-1}$), the 375 cm$^{-1}$ band exhibits a more pronounced feature in our experimental spectrum and is therefore tentatively assigned to the calculated mode at 370 cm$^{-1}$. Compared to prior research, this work found several new bands for 1-CNN and provides the first measurement of several overtones or combination bands in the 1600–2000 cm$^{-1}$ range based on theoretical calculations.

**Table 1.** Comparison of the calculated (B3LYP/N07D) and experimental vibrational frequencies of 1-CNN. The numbers in parentheses indicate the difference between the experimental and harmonic frequencies. Frequencies labeled as Harm. and Obs. are in unit of cm$^{-1}$, while the harmonic intensities (Harm. Int.) are in unit of km·mol$^{-1}$. For additional information on other fundamental, overtone, and combination bands, see the Appendix A.

| | Harm. | Harm. Int. | Obs. | Obs. Int. | Assignment |
|---|---|---|---|---|---|
| $\nu_5$ | 351 | 2.34 | 355 (4) | 0.21 | CC in-plane bending |
| $\nu_8$ | 451 | 14.31 | 447 (-4) | 1.47 | CC out-of-plane bending |
| $\nu_9$ | 466 | 1.66 | 466 (0) | 0.40 | CC out-of-plane bending |
| $\nu_{11}$ | 538 | 1.18 | 542 (4) | 0.32 | CC in-plane bending |
| $\nu_{12}$ | 573 | 1.74 | 567 (-6) | 0.50 | CC out-of-plane bending |
| $\nu_{15}$ | 690 | 3.23 | 685 (-5) | 0.66 | CC in-plane bending; CC stretch |
| $\nu_{17}$ | 764 | 42.49 | 769 (5) | 2.58 | CH/CC out-of-plane bending |
| $\nu_{19}$ | 794 | 63.30 | 798 (4) | 2.28 | CH/CC out-of-plane bending |
| $\nu_{40}$ | 1514 | 11.52 | 1515 (1) | 0.58 | CH/CC in-plane bending; CC stretch |

### *3.2. 2-CNN infrared spectrum*

The experimental and theoretical infrared spectra of 2-CNN are presented in Fig. 1(b), with the assignments of selected fundamental vibrational modes of relatively strong intensity listed in Table 2. More comprehensive assignments of fundamental vibrational modes are provided in Table A.3. All assigned vibrational modes are explicitly labeled in Fig. 2. Similar as in the case of 1-CNN, in the 350–1650 cm$^{-1}$ region, the harmonic spectrum exhibits good agreement with the experiment, with the harmonic frequencies scaled by a factor of 0.976. Several strong

characteristic peaks are observed at 470, 740, 810, 854, 898, 954, 1159, 1268, 1363, and 1504 $cm^{-1}$. The peak at 470 $cm^{-1}$ is attributed to a CC out-of-plane bending mode, the 740 $cm^{-1}$ band is mainly contributed by a CH out-of-plane bending mode, both the 810 and 854 $cm^{-1}$ bands originate from the combined contributions of CC and CH out-of-plane bending modes, the 898 $cm^{-1}$ band is mainly dominated by a CH out-of-plane bending mode, and the 954 $cm^{-1}$ band is assigned to a CC in-plane bending mode. The bands at 1159, 1268, 1363, and 1504 $cm^{-1}$ are all associated with CH in-plane bending and C–C stretching modes.

Similar to the case of 1-CNN, not all calculated fundamental features are observed in the experimental spectrum. For example, a feature is predicted at 752 $cm^{-1}$ ($\nu_{17}$) with a vibrational frequency close to that of $\nu_{16}$, yet no distinct independent feature was resolved experimentally. It is hypothesized that this feature likely overlaps with the predicted peak at 740 $cm^{-1}$ ($\nu_{16}$) or 765 $cm^{-1}$ ($\nu_{18}$) and remains unresolved at the current experimental resolution. The predicted $\nu_{21}$ (876 $cm^{-1}$) also does not manifest independently. Regarding $\nu_{23}$ and $\nu_{24}$, their calculated frequencies suggest they may merge into the single peak at 954 $cm^{-1}$ in the experimental spectrum. Likewise, the two closely spaced modes $\nu_{30}$, and $\nu_{31}$ likely overlap to form the feature observed at 1159 $cm^{-1}$. Detailed information for all the vibrational modes mentioned above is provided in Table A.3.

**Table 2.** Comparison of the calculated (B3LYP/N07D) and experimental vibrational frequencies of 2-CNN. The numbers in parentheses indicate the difference between the experimental and harmonic frequencies. Frequencies labeled as Harm. and Obs. Are in unit of $cm^{-1}$, while the harmonic intensities (Harm. Int.) are in unit of $km \cdot mol^{-1}$.

| | Harm. | Harm. Int. | Obs. | Obs. Int. | Assignment |
|---|---|---|---|---|---|
| $\nu_5$ | 349 | 2.40 | 351 (2) | 0.42 | CC in-plane bending |
| $\nu_9$ | 473 | 20.05 | 470 (-3) | 0.97 | CC out-of-plane bending |
| $\nu_{16}$ | 740 | 33.60 | 740 (0) | 1.24 | CH out-of-plane bending |
| $\nu_{19}$ | 812 | 37.18 | 810 (-2) | 1.14 | CH/CC out-of-plane bending |
| $\nu_{20}$ | 855 | 16.30 | 854 (-1) | 0.95 | CH/CC out-of-plane bending |
| $\nu_{22}$ | 895 | 13.73 | 898 (3) | 0.79 | CH out-of-plane bending |
| $\nu_{23}$ | 946 | 2.12 | 954 | 0.63 | CH out-of-plane bending |
| $\nu_{24}$ | 950 | 3.47 | | | CC in-plane bending; CC stretch |
| $\nu_{27}$ | 1021 | 2.68 | 1017 (-4) | 0.32 | CH in-plane bending; CC stretch |
| $\nu_{30}$ | 1153 | 1.42 | 1159 | 0.47 | CH in-plane bending; CC stretch |
| $\nu_{31}$ | 1160 | 4.35 | | | CH in-plane bending; CC stretch |
| $\nu_{34}$ | 1262 | 7.06 | 1268 (6) | 0.42 | CH in-plane bending; CC stretch |
| $\nu_{36}$ | 1375 | 5.39 | 1363 (-12) | 0.33 | CH in-plane bending; CC stretch |
| $\nu_{40}$ | 1508 | 5.84 | 1504 (-4) | 0.41 | CH/CC in-plane bending; CC stretch |

In the 1600–2000 $cm^{-1}$ region, anharmonic calculations provide a more reliable interpretation of the experimental spectrum. The corresponding assignments are summarized in Table A.4 of Appendix A. The black arrows in Fig. 2 indicate the experimentally observed bands attributed to these anharmonic transitions. Their agreement with the anharmonic calculations further supports the reliability of these assignments. A comparison of our results with previous assignments based on FTIR spectroscopy reveals discrepancies in several mode assignments. We assign the 810 $cm^{-1}$ band ($\nu_{18}$) to CH and CC in-plane bending mode, which differs from the CH out-of-plane bending assignment in previous FTIR work (Rawat et al. 2025). Our assignment is supported by the previous investigation of 2-CNN in low-temperature ice phases. Further discussion regarding the different vibrational modes is provided in the Appendix A. A comparison with the recently reported gas-phase FTIR spectrum of 2-CNN (Bentley et al. 2026) also shows good agreement for the commonly observed low-frequency modes. The bands observed in this work at 351, 470, 543, 616, and 638 $cm^{-1}$ correspond closely to the reported features at 351, 473, 543, 619, and 641 $cm^{-1}$, respectively, with deviations generally within 3 $cm^{-1}$. Compared to previous studies, this work identifies several new bands for 2-CNN and provides the first identification of several overtones or combination bands in the 1600–2000 $cm^{-1}$ range.

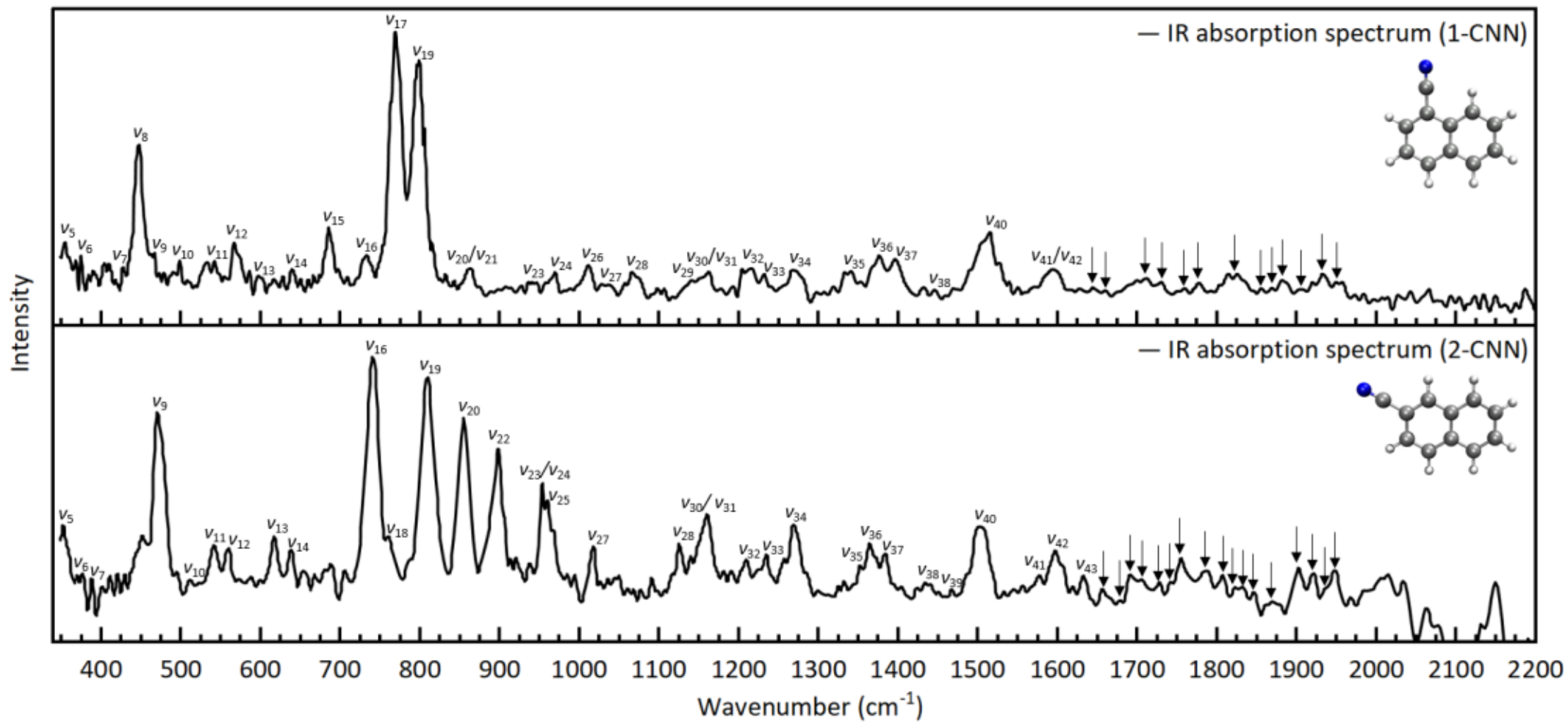

**Fig. 2.** Vibrational assignments of characteristic peaks for 1-CNN and 2-CNN. Black arrows indicate the presence of combination and overtone transitions. The chemical structures of CNN isomers are included.

### *3.3. IR emission spectra of CNNs*

Infrared emissions from PAHs in the interstellar medium are governed by a non-equilibrium stochastic heating mechanism. Since small PAHs possess a limited number of vibrational degrees of freedom and low heat capacities, absorption of a single UV or visible photon can substantially increase their internal energy. This energy is rapidly converted into vibrational excitation through non-radiative relaxation, resulting in a transient vibrational temperature that may reach several thousand Kelvin. In weak to moderate radiation fields, the time interval between successive photon absorption events generally exceeds the infrared cooling time, so that the molecule cools to a low-energy vibrational state before the next absorption occurs. This condition defines the single-photon heating limit, under which individual absorption–emission cycles are independent. The resulting PAH infrared emission spectrum can therefore be represented as a weighted linear superposition of spectra corresponding to different absorbed photon energies.

To investigate the connection between AIBs and CNNs, we simulated the infrared emission spectra of 1-CNN and 2-CNN based on their experimental spectra, as shown in Fig. 3(b) and Fig. 3(c). This study employ the PAH emission model based on the Single Photon Approximation (SPA) developed by Richie & Hensley (2026). By utilizing pre-calculated basis spectra excited at various photon energies and applying the principles of energy conservation and linear superposition, this framework simulates infrared emission during the stochastic heating process. Compared to traditional transition matrix methods, the SPA framework significantly reduces computational cost while maintaining high physical accuracy, with errors below 10% for radiation field intensities $U<100$. The computational procedure begins by constructing a library of basis spectra for specific grain sizes. For each absorption wavelength, $\lambda_{abs}$, the radiative cooling equation is solved by numerical integration. The resulting basis spectra are then scaled and weighted according to the spectral energy distribution of the input radiation field. The final synthetic emission spectrum is obtained by summing the weighted basis spectra and can be directly compared with astronomical observations, such as those obtained with the Mid-Infrared Instrument (MIRI) onboard the James Webb Space Telescope (JWST).

The emission model was first validated by reproducing a previously published spectrum using the absorption cross sections reported in that study, yielding results consistent with the published data (Li et al. 2024). The detailed comparison between the reproduced and published spectra is presented in Appendix B. For the molecules investigated here, however, the experimental infrared spectra were converted to approximate absolute absorption cross sections through a global normalization procedure. The normalization constant was adopted only to place the spectra on a representative PAH absorption scale and should not be regarded as a molecule-specific absolute calibration. Although this treatment is sufficient for identifying the principal vibrational features, the absolute intensity scale and the resulting continuum level are less well constrained. Therefore, the slowly varying background is not interpreted quantitatively, and the following discussion is restricted to the positions of the characteristic vibrational bands.

The spectral resolution of the SPA simulated emission spectra is inherently limited by that of the input IR absorption spectra used in the model. The IR absorption spectra previously used by Li et al. for emission modeling were limited by their relatively low spectral resolution, which led to substantial overlap among adjacent spectral features. In contrast, the gas-phase IR absorption spectra reported in this study exhibit considerably higher spectral resolution. As a result, the simulated emission spectra of 1-CNN and 2-CNN reveal several features that could not be resolved in the previous low-resolution spectra. For 1-CNN, these newly resolved features include the CC in-plane bending mode at 20.2 µm, the CC out-of-plane bending mode at 15.6 µm, the CH out-of-plane bending mode at 13.6 µm, and the coupled CH/CC out-of-plane bending mode at 11.6 µm. In addition, some of these newly resolved features are assigned to overtones and combination bands that have not been previously reported. These additional spectral markers may provide more detailed templates for future comparisons with JWST observational data.

## 4. Astrophysical relevance

Orion Bar is an important astrophysical object for studying the mid-infrared emissions of PAHs. With the unprecedented spatial resolution and sensitivity of JWST, several recent studies have provided refined analyses of the spectral characteristics of AIBs and their correspondence with PAH vibrational modes (Chown et al. 2024; Pasquini et al. 2024; Peeters et al. 2024). As a prototypical and extensively studied photodissociation region (PDR), Orion Bar offers new insights into the physicochemical evolution and molecular structures of PAHs in the interstellar medium.

In the 6–9 µm range, the 6.2 µm band is one of the main AIB features and is generally attributed to CC stretching vibrations in aromatic structures, which is more prominent in cationic PAHs (Allamandola et al. 1999). Our simulated emission spectra of 1-CNN and 2-CNN both exhibit clear features near 6.2–6.3 μm, which can be assigned to CC stretching vibrations. In addition to the 6.2 μm main band, 1-CNN shows a distinct feature near 6.6 μm, and 2-CNN shows a similar feature near 6.7 μm. These features can be attributed to coupled CH in-plane bending and CC stretching vibrations. Furthermore, both 1-CNN and 2-CNN display weak vibrational features near 7.0 μm, which are likely associated with CH in-plane bending and CC stretching vibrations. Correspondingly, weak AIB features have been reported at 6.94 and 6.64 μm (Arnoult et al. 2000; Chown et al. 2024), where the former is commonly assigned to a CH in-plane bending vibration and the latter is considered to arise from mixed CH in-plane bending and CC stretching modes. Thus, several emission features of 1-CNN and 2-CNN in the 6–7 μm range show a consistency with the observational weak AIB features suggesting that CNNs may contribute to some of the astronomical emission features in this wavelength range.

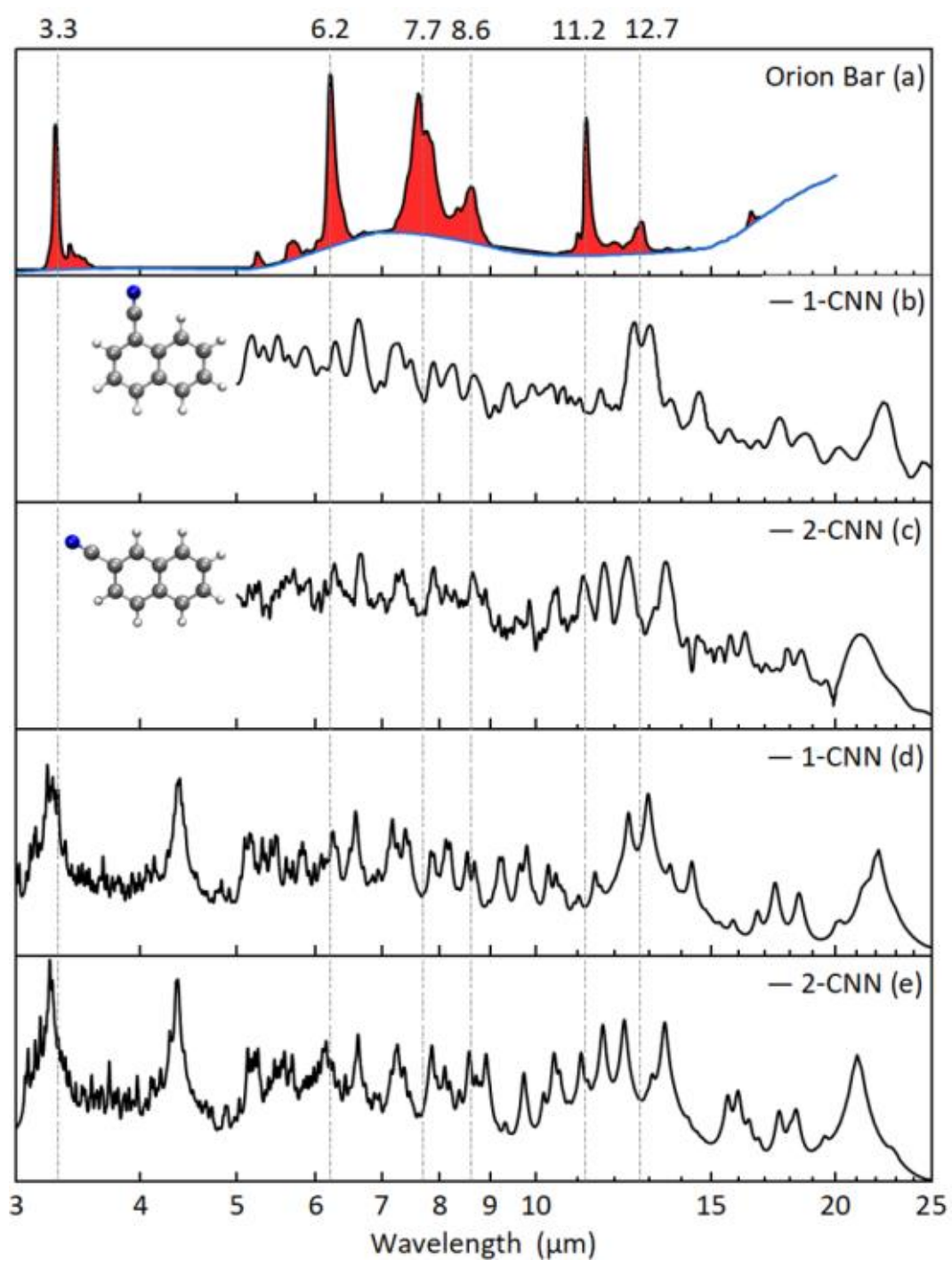

**Fig. 3.** Comparison of the Orion Bar AIB spectrum with simulated infrared emission spectra of 1-CNN and 2-CNN. (a) JWST Orion Bar atomic PDR template spectrum (Chown et al. 2024), where the red-shaded regions represent AIB emission and the blue curve indicates the underlying continuum; (b, c) Simulated IR emission spectra of 1-CNN and 2-CNN derived from the experimental infrared absorption spectra, respectively. (d, e) Simulated IR emission spectra of 1-CNN and 2-CNN obtained from anharmonic vibrational calculations, respectively. The black vertical dashed lines mark the positions of the major AIB features at 3.3, 6.2, 7.7, 8.6, 11.2, and 12.7 μm. The chemical structures of CNN isomers are included.

There is evidence that the 7.7 μm AIB feature is not a single band, but consists of multiple subfeatures(Cohen et al. 1989; Peeters et al. 2002). JWST spectra show that the main component of this band is at 7.626 μm, accompanied by bands of moderate intensity at 7.8 and 7.85 μm (Chown et al. 2024). Previous studies generally concluded that the features in this wavelength region mainly arise from mixed CC stretching and CH in-plane bending vibrations. In comparison, our simulated emission spectra of 1-CNN and 2-CNN show two features near 7.3 and 7.9 μm, respectively, and these features can be assigned to coupled CC stretching and CH in-plane bending vibrations. While consistent with the conventional vibrational assignment of the 7.7 μm AIB, the peak positions of 1-CNN and 2-CNN show a measurable offset from astronomical observations, indicating that these molecules provide useful reference for understanding the vibrational character of this spectral region, although further investigation is needed to fully account for the observed AIBs.

The 8.6 μm AIB is generally attributed to the CH in-plane bending vibrations of PAHs, with some contribution from CC stretching vibrations(Bauschlicher Jr et al. 2008). Our simulated spectra show that both 1-CNN and 2-CNN exhibit distinct features at 8.6 μm, which can be assigned to mixed CH in-plane bending and CC stretching modes. This is consistent with the assignment of the 8.6 μm AIB. In addition to the 8.6 μm feature, the simulated emission spectrum of 2-CNN also shows a weaker feature at 8.9 μm, which has same vibrational assignment as the 8.6 μm feature. It is noted that a weak AIB feature at 8.9 μm has been tentatively detected in astronomical observations (Chown et al. 2024). Overall, compared to the 7.7 μm region, 1-CNN and 2-CNN show better agreement with the observed AIB features in the 8.6 μm range.

Previous theoretical and experimental studies have attributed the 10–15 μm AIBs primarily to PAHs CH out-of-plane bending modes (Hony et al. 2001), and both theoretical calculations and laboratory infrared spectroscopy suggest that the CH out-of-plane bending vibrations of PAHs can account for the major AIB features in this spectral region (Oomens et al. 2003; Zhen et al. 2018). For example, Hony et al. attributed the 10.6 μm AIB feature to the solo modes of cationic PAHs (Hony et al. 2001), while Bauschlicher et al. suggested that the AIB bands at 9.5, 10.2, and 10.6 μm are more likely associated with the CC stretching vibrations and benzene ring deformations involving quartet hydrogens (Bauschlicher Jr et al. 2009). In our simulated emission spectrum, 1-CNN exhibits weak feature near 10.6 μm, which can be assigned to the CH out-of-plane bending vibration. For 2-CNN, several vibrational modes are found near 10.5 μm, most of which correspond to CH out-of-plane bending vibrations with one mode associated with ring deformations. Hence, both 1-CNN and 2-CNN show multiple features related to CH out-of-plane bending vibrations in the 10–11 μm range, suggesting that the AIBs in this region may contain contributions from neutral aromatic molecules in addition to those commonly attributed to cationic PAHs.

The 11.0 μm AIB band is generally assigned to the solo CH out-of-plane vibration of PAH cations, the 11.207 μm band to the solo CH vibration of neutral PAHs, and the 11.25 μm band may be associated with the solo CH out-of-plane mode of PAH clusters (Hudgins & Allamandola 1999a; Bauschlicher Jr et al. 2008). In our simulated spectra, the 11.6 μm feature of 1-CNN can be attributed to the out-of-plane bending mode of CH and CC groups. The emission spectrum of 2-CNN shows a prominent peak at 11.1 μm arising from the vibrational mode of CH group, consistent with the assignment of the corresponding AIB feature observed in the Orion Bar (Khan et al. 2025).

In the 12–13 μm spectral range, the 12.0 μm AIB is usually attributed to the duo CH out-of-plane modes, and the 12.7 μm band is commonly considered to include contributions from both duo and trio CH out-of-plane modes (Hony et al. 2001; Bauschlicher Jr et al. 2009). Our simulated emission spectrum of 1-CNN shows strong features near 12.6 and 13.0 μm, mainly arising from CH out-of-plane bending vibrations. The emission spectrum of 2-CNN shows no obvious feature at 12.0 μm, but displays two CH out-of-plane bending features associated with CH groups near 11.7 and 12.3 μm. Given the spectral resolution of astronomical observations, these two features may overlap and appear observationally as a single broader feature at 12.0 μm.

The 13.5 μm AIB is generally considered to arise from the out-of-plane bending vibration of quartet CH groups of PAHs (Hony et al. 2001; Bauschlicher Jr et al. 2009). In our simulated spectra, 1-CNN exhibits a feature near 13.6 μm from the out-of-plane bending vibration of CH groups, and 2-CNN also shows a peak near 13.5 μm associated with CH groups. These results are consistent with the traditional assignment of the 13.5 μm AIB.

For the AIBs in the 14 μm region, previous studies suggested that the 14.2 μm band is more likely related to a CC bending mode of positively charged species (Hony et al. 2001; Khan et al. 2025). In our simulated spectra, 1-CNN shows a clear feature at 14.6 μm, which can be attributed to the bending and stretching vibrations of the benzene ring. Although the 14.6 μm feature of 1-CNN involves vibrational motions similar to those proposed for the 14.2 μm AIB, its frequency does not coincide with the astronomical features, suggesting that the origins of the AIBs in this spectral region still require further constraints.

The plateau features of the AIBs in the 15–20 μm range are attributed to CC skeleton vibrational modes, with the 16.4 μm feature being most prominent (Chown et al. 2024). In our simulations, the 1-CNN emission spectrum exhibits weak features at 15.6, 16.7, 17.6, and 18.6 μm, and 2-CNN also shows weak features at 15.7, 16.2, 17.9, and 18.5 μm. These simulated features correspond to the CC bending vibrations of the benzene rings. Given that the AIB features in this spectral region are generally weak and mainly reflect the PAH CC skeleton vibrations, they do not provide distinct signatures that could allow a direct connection between the observed AIBs and specific PAH molecules.

In addition, we simulated the infrared emission spectra of 1-CNN and 2-CNN below 5 μm based on the anharmonic calculations, as shown in Fig. 3(d) and Fig. 3(e). It should be noted that the CN stretching region was not covered experimentally in this work. This spectral region is of particular interest, as CN stretching vibrations have recently attracted attention as potential diagnostic features for identifying cyano-PAHs in JWST observations (Esposito et al. 2024; Esposito et al. 2025). These calculated spectra allow a comparative analysis with the astronomical AIB features in the 3–6 μm range. Recent JWST observations revealed a strong 3.29 μm AIB in the 3 μm region, along with weaker bands at 3.25, 3.40, 3.46, 3.52, and 3.56 μm superimposed on a broad plateau, and these features are generally attributed to aromatic and aliphatic CH stretching vibrations (Geballe et al. 1989; Tokunaga et al. 1991; Sloan et al. 1997; Van Diedenhoven et al. 2004). In addition to the known AIB features, two broad emission bands were reported near 3.8 and 4.5 μm (Peeters et al. 2024). In the 4 μm region, previous studies detected features near 4.4 and 4.6 μm in the Orion Bar, which were attributed to deuterated PAHs (Peeters et al. 2004; Onaka et al. 2022). Previous studies have suggested that the broad bands at 3.8 and 4.35 μm may overlap with nitrile stretching bands near 4.38 and 4.52 μm, as well as with CD stretching vibrations of deuterated PAHs in the 4.3–4.5 μm region (Hudgins et al. 2004; Allamandola et al. 2021; Peeters et al. 2024). Therefore, although the CN stretching modes of 1-CNN and 2-CNN were not experimentally measured in this work, the theoretical results provide valuable reference data for assessing the possibility that these broad bands may originate from cyano-substituted aromatic molecules or deuterated aromatic species.

The simulated emission spectra of 1-CNN and 2-CNN both exhibit a strong and broad feature near 3.2–3.3 μm, which can be attributed to the CH stretching vibrations. No feature is observed around 3.4 μm, consistent with the conventional assignment of the 3.3 μm AIB. In addition to the main peak, a weaker feature appears near 3.7 μm in both 1-CNN and 2-CNN spectra. According to the anharmonic calculations, this feature corresponds to a mixed mode of CH in-plane bending and benzene skeletal vibrations. Besides, both 1-CNN and 2-CNN exhibit relatively strong features near 4.4 μm due to CN stretching vibrations. Considering the proximity of this band to the 4.35–4.5 μm broad emission structures reported in literature, these results indicate that cyano-substituted aromatic molecules may contribute significantly to the spectral features in this region.

In the 5–6 μm region, previous observations identified two moderately weak AIB features at 5.25 and 5.75 μm (Allamandola et al. 1989a; Boersma et al. 2009). The PAH spectra in this spectral region display weak combination bands associated with the CH out-of-plane vibrations. The simulated emission spectrum of 1-CNN shows features centered at 5.2 and 5.8 μm, and 2-CNN exhibits features centered at 5.2 and 5.7 μm. These features can be assigned to the combination bands involving CH out-of-plane bending vibrations. Comparison of these results suggests that the simulated emission spectra of 1-CNN and 2-CNN could reproduce the weak AIB features in the 5–6 μm range.

A systematic comparison between the simulated emission spectra of 1-CNN and 2-CNN and the astronomical AIBs suggests that CNNs exhibit consistency with the AIB features across multiple spectral regions. In the 6.2 μm band, both 1-CNN and 2-CNN display CC stretching features corresponding to the classical AIB band. Within the 8.6–8.9 μm range, the simulated spectra match the AIBs, corresponding to mixed CH in-plane bending and CC stretching motions. For the 10–15 μm region, the CNN simulated spectra exhibit prominent features at 11.1, 12.6, and 13.5 μm, corresponding respectively to the solo, trio, and quartet CH out-of-plane bending modes. This is in good agreement with the classical attribution of AIBs in the Orion Bar. Notably, the solo CH out-of-plane bending feature of 2-CNN at 11.1 μm is highly comparable to the strongest 11.2 μm AIB observed astronomically, while the pronounced trio CH out-of-plane bending feature of 1-CNN at 12.6 μm further supports the inherent correspondence between the peripheral hydrogen arrangement in CNN and that of interstellar aromatic carriers.

In addition, in the 3–5 μm range, the simulated emission spectra of 1-CNN and 2-CNN exhibit a strong feature at 3.2–3.3 μm, consistent with aromatic CH stretching vibrations, and a pronounced response near 4.4 μm, attributable to CN stretching vibrations. These results suggest that CNNs not only reproduce the classical aromatic vibrational features of AIBs, but also provides complementary molecular diagnostic information in the nitrile-related spectral region. Although some discrepancies exist in certain bands, for example near 7.7 and 14 μm, including slight peak shifts or complex mode assignments, these differences primarily reflect the multi-component nature and intrinsic complexity of AIB carriers, and do not negate the potential connection between CNNs and the AIBs.

## 5. Conclusions

In summary, the gas-phase infrared spectra of neutral 1-CNN and 2-CNN were measured for the first time using ion-dip spectroscopy and assigned with the aid of harmonic and anharmonic quantum chemical calculations. The harmonic calculations provided reliable assignments for fundamental vibrational modes, and the VPT2 anharmonic calculations enabled the interpretation of high-wavenumber features involving overtone and combination bands. Compared with previous condensed-phase or lower-resolution studies, the present gas-phase spectra provide more detailed vibrational information for both CNN isomers.

Based on the experimental infrared spectra, infrared emission spectra of 1-CNN and 2-CNN were simulated and compared with astronomical AIB features. The results show that both molecules reproduce or closely approach several key AIB regions. Their simulated spectra exhibit aromatic CH stretching features near 3.2–3.3 μm, CC stretching features near the classical 6.2 μm AIB, and mixed CH in-plane bending/CC stretching modes in the 8.6–8.9 μm range. In the 10–15 μm region, 2-CNN shows a clear CH out-of-plane bending feature near 11.1 μm, while 1-CNN exhibits prominent CH out-of-plane bending features near 12.6 μm. These correspondences suggest that the peripheral hydrogen structures of CNN molecules are compatible with those inferred for interstellar aromatic AIB carriers.

Although discrepancies remain in some regions, especially near 7.7 μm and 14 μm, they are consistent with the multi-component nature of AIB carriers. Overall, the present results suggest that 1-CNN and 2-CNN are promising molecular contributors to AIBs and merit further investigation through high-resolution astronomical observations.

## Data availability

Appendix is available at 10.5281/zenodo.21817859

*Acknowledgements.* This work was supported by the National Natural Science Foundation of China (No.223B2306), the Natural Science Basic Research Program of Shaanxi Province (2025JC-JCQN-043), the Postdoctoral Fellowship Program of CPSF (GZC20250726), and the Fundamental Research Funds for Central Universities, China. The authors gratefully acknowledge the Nederlandse Organisatie voor Wetenschappelijk Onderzoek (NWO) for the support of the HFML-FELIX Institute.

## Appendix A: Experimental spectra of 1-CNN and 2-CNN.

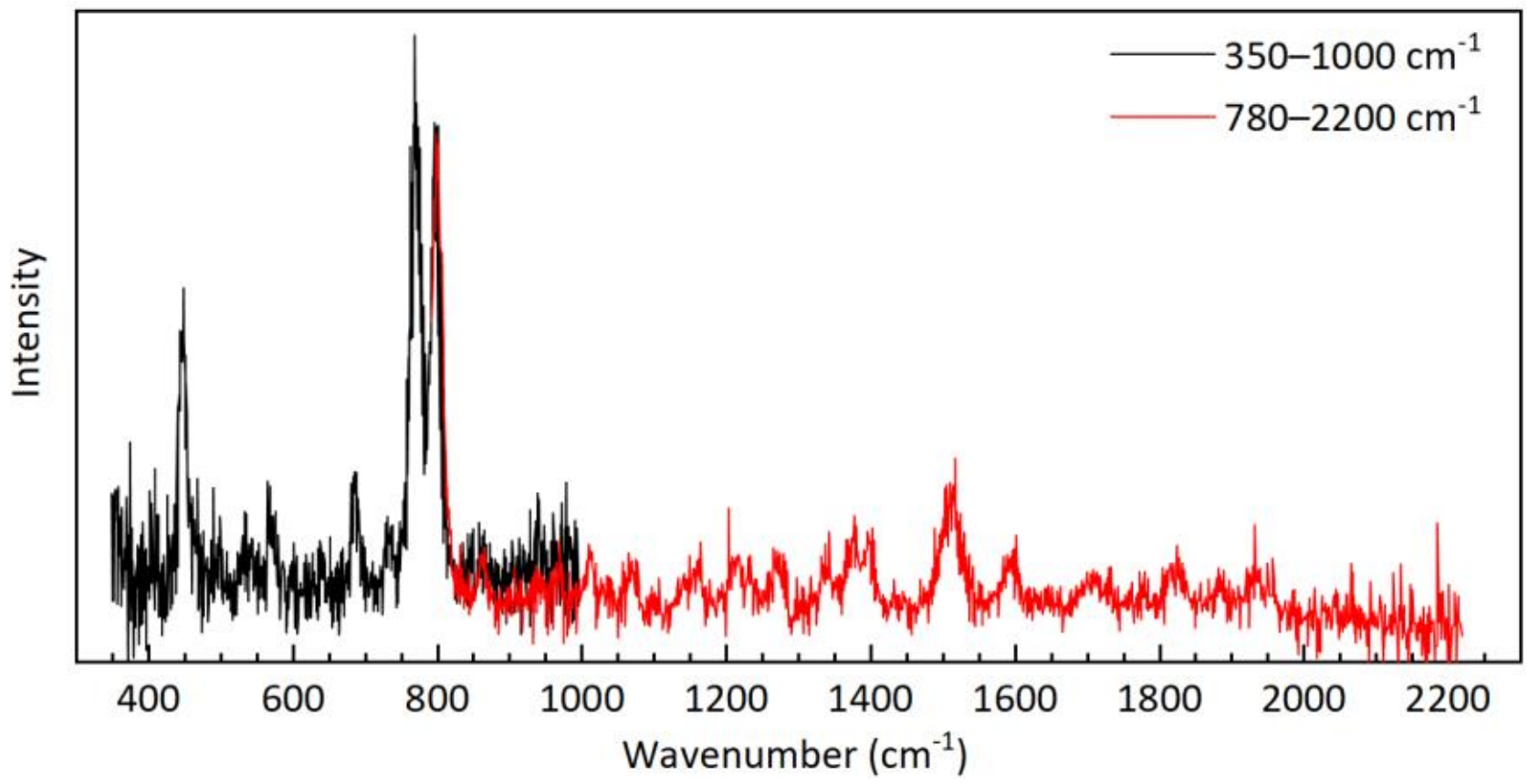

**Fig. A.1.** Experimental spectra of 1-CNN. The final experimental spectrum is a composite of measurements from two spectral ranges: the black trace represents the 350–1000 cm$^{-1}$ region, and the red trace represents the 780–2200 cm$^{-1}$ region.

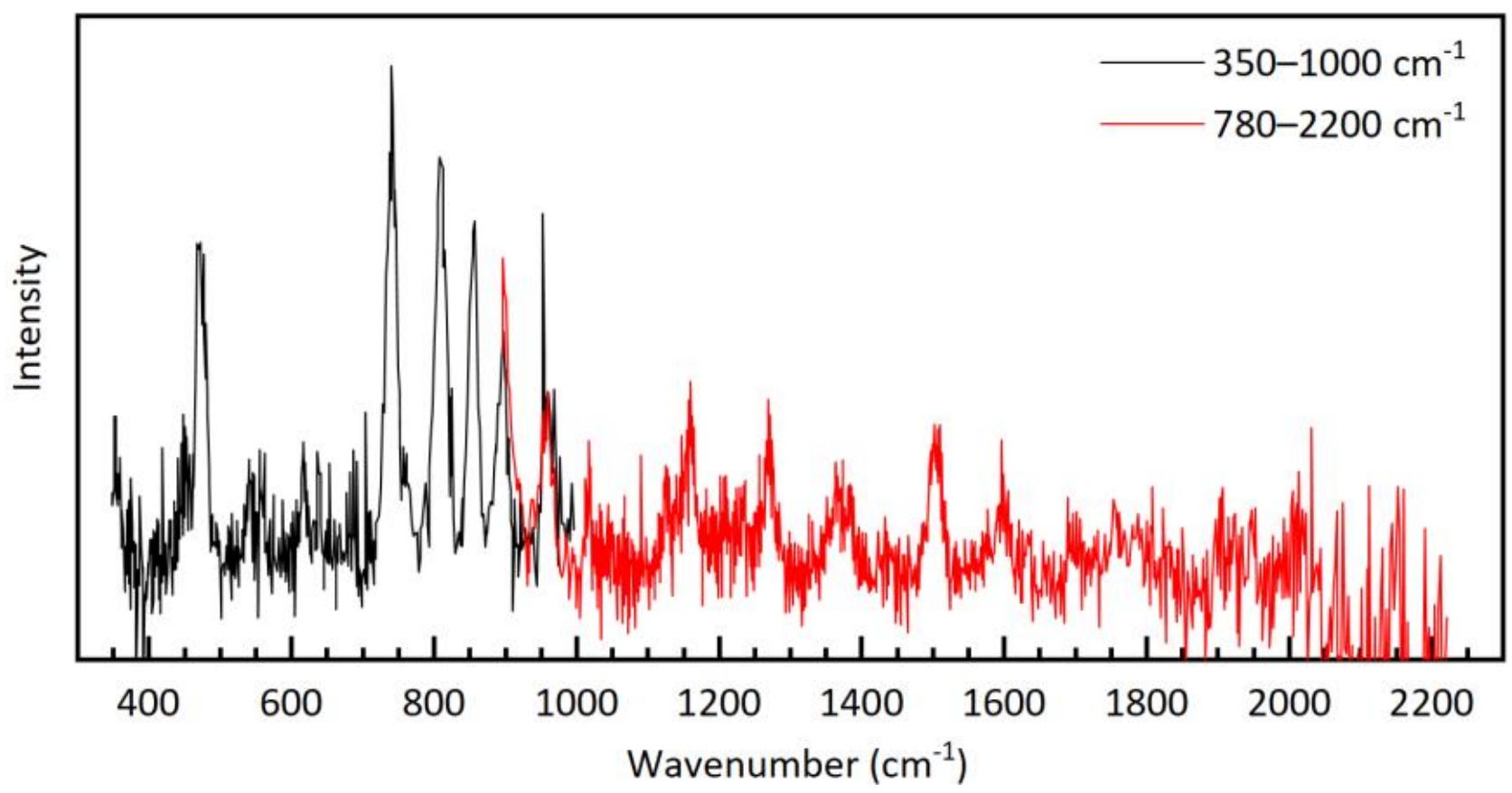

**Fig. A.2.** Experimental spectra of 2-CNN. The final experimental spectrum is a composite of measurements from two spectral ranges: the black trace represents the 350–1000 cm$^{-1}$ region, and the red trace represents the 780–2200 cm$^{-1}$ region.

**Table A.1.** Experimental and B3LYP/N07D calculated vibrational frequencies and the vibrational assignments for 1-CNN. The numbers in parentheses indicate the difference between the experimental and harmonic frequencies. Frequencies labeled as Harm. and Obs. are in unit of $cm^{-1}$, while the harmonic intensities (Harm. Int.) are in unit of $km \cdot mol^{-1}$.

| | Harm. | Harm. Int. | Obs. | Obs. Int. | Assignment | Cal.[a] | FTIR[a] | Assignment[a] |
|---|---|---|---|---|---|---|---|---|
| $\nu_5$ | 351 | 2.34 | 355 (4) | 0.21 | α(CC) | 360 | 356 | α(CC) |
| $\nu_6$ | 370 | 0.95 | 375 (5) | 0.38 | τ(CC) | 375 | 372 | τ(CC) |
| $\nu_7$ | 430 | 0.21 | 428 (-2) | 0.27 | α(CC) | 436 | 434 | α(CC) |
| $\nu_8$ | 451 | 14.31 | 447 (-4) | 1.47 | τ(CC) | 459 | 450 | τ(CC) |
| $\nu_9$ | 466 | 1.66 | 466 (0) | 0.40 | τ(CC) | 470 | 465 | τ(CC) |
| $\nu_{10}$ | 490 | 0.22 | 498 (8) | 0.22 | α(CC) | 500 | 493 | α(CC) |
| $\nu_{11}$ | 538 | 1.18 | 542 (4) | 0.32 | α(CC) | 545 | 539 | α(CC) |
| $\nu_{12}$ | 573 | 1.74 | 567 (-6) | 0.50 | τ(CC) | 585 | 573 | τ(CC) |
| $\nu_{13}$ | 590 | 0.33 | 597 (7) | 0.18 | α(CC) | 606 | 594 | α(CC) |
| $\nu_{14}$ | 631 | 0.06 | 639 (8) | 0.25 | τ(CC) | 644 | 631 | τ(CC) |
| $\nu_{15}$ | 690 | 3.23 | 685 (-5) | 0.66 | α(CC), R(CC) | 696 | 692 | α(CC), R(CC) |
| $\nu_{16}$ | 728 | 0.74 | 733 (5) | 0.39 | ε(CH) | 739 | 748 | ε(CH) |
| $\nu_{17}$ | 764 | 42.49 | 769 (5) | 2.58 | ε(CH), τ(CC) | 776 | 773 | ε(CH), α(CC) |
| $\nu_{18}$ | 787 | 0.56 | — | — | α(CC) | 797 | 791 | α(CC) |
| $\nu_{19}$ | 794 | 63.30 | 798 (4) | 2.28 | ε(CH), τ(CC) | 805 | 800 | τ(CC), β(CH) |
| $\nu_{20}$ | 852 | 0.29 | 861 | 0.26 | α(CC), R(CC) | 860 | 856 | α(CC) |
| $\nu_{21}$ | 859 | 0.99 | | | ε(CH), τ(CC) | 870 | 868 | ε(CH), τ(CC) |
| $\nu_{22}$ | 912 | 0.14 | — | — | ε(CH) | 922 | 923 | ε(CH) |
| $\nu_{23}$ | 950 | 0.80 | 946 (-4) | 0.11 | ε(CH) | 964 | 954 | ε(CH) |
| $\nu_{24}$ | 968 | 1.57 | 969 (1) | 0.29 | ε(CH) | 980 | 982 | ε(CH) |
| $\nu_{25}$ | 977 | 0.01 | — | — | ε(CH) | 987 | 987 | ε(CH) |
| $\nu_{26}$ | 1014 | 3.25 | 1011 (-3) | 0.29 | β(CH), α(CC), R(CC) | 1018 | 1013 | β(CH), R(CC) |
| $\nu_{27}$ | 1030 | 1.58 | 1032 (2) | 0.10 | β(CH), α(CC), R(CC) | 1038 | 1038 | β(CH), α(CC) |
| $\nu_{28}$ | 1073 | 3.35 | 1069 (-4) | 0.20 | β(CH), R(CC) | 1080 | 1072 | β(CH), R(CC) |
| $\nu_{29}$ | 1143 | 1.43 | 1141 (-2) | 0.14 | β(CH), α(CC), R(CC) | 1152 | 1141 | β(CH), α(CC), R(CC) |
| $\nu_{30}$ | 1159 | 1.30 | 1162 | 0.20 | β(CH), R(CC) | 1171 | 1171 | β(CH), R(CC) |
| $\nu_{31}$ | 1166 | 0.58 | | | β(CH), R(CC) | 1178 | — | R(CC) |
| $\nu_{32}$ | 1217 | 3.89 | 1215 (-2) | 0.26 | β(CH), R(CC) | 1215 | 1213 | β(CH), R(CC) |
| $\nu_{33}$ | 1226 | 3.25 | 1231 (5) | 0.20 | β(CH), R(CC) | 1229 | 1230 | β(CH), α(CC) |
| $\nu_{34}$ | 1264 | 2.60 | 1267 (3) | 0.24 | β(CH), α(CC), R(CC) | 1276 | 1270 | β(CH), α(CC), R(CC) |
| $\nu_{35}$ | 1345 | 6.84 | 1341 (-4) | 0.23 | β(CH), α(CC), R(CC) | 1349 | 1343 | β(CH), α(CC), R(CC) |
| $\nu_{36}$ | 1378 | 6.65 | 1377 (-1) | 0.38 | α(CC), R(CC) | 1360 | 1375 | α(CC), R(CC) |
| $\nu_{37}$ | 1394 | 2.93 | 1395 (1) | 0.35 | β(CH), R(CC) | 1398 | 1393 | β(CH) |
| $\nu_{38}$ | 1439 | 0.38 | 1446 (7) | 0.05 | β(CH), α(CC), R(CC) | 1435 | 1437 | β(CH), α(CC), R(CC) |
| $\nu_{39}$ | 1460 | 0.16 | — | — | β(CH), α(CC), R(CC) | 1466 | 1463 | β(CH), α(CC), R(CC) |
| $\nu_{40}$ | 1514 | 11.52 | 1515 (1) | 0.58 | β(CH), α(CC), R(CC) | 1517 | 1513 | β(CH), α(CC), R(CC) |

**Table A.1.** continued.

| | Harm. | Harm. Int. | Obs. | Obs. Int. | Assignment | Cal.[a] | FTIR [a] | Assignment [a] |
|---|---|---|---|---|---|---|---|---|
| $\nu_{41}$ | 1585 | 2.73 | 1596 | 0.25 | β(CH), α(CC), R(CC) | 1578 | 1577 | β(CH), α(CC), R(CC) |
| $\nu_{42}$ | 1597 | 3.01 | | | β(CH), α(CC), R(CC) | 1591 | 1591 | α(CC), R(CC) |
| $\nu_{43}$ | 1630 | 0.59 | 1628 (-2) | 0.06 | β(CH), α(CC), R(CC) | 1621 | 1623 | α(CC), R(CC) |

**Notes:** R(CC): CC stretching; β(CH) and α(CC): CH and CC in-plane bending; ε(CH) and τ(CC): CH and CC out-of-plane bending; [a]These data were taken from Rawat et al. (2025)

**Table A.2.** Experimental and B3LYP/N07D calculated vibrational frequencies and the vibrational assignments (overtones and combination bands) for 1-CNN. The numbers in parentheses indicate the difference between the experimental and anharmonic frequencies. Frequencies labeled as Anharm. and Obs. are in unit of $cm^{-1}$, while the anharmonic intensities (Anharm. Int.) are in unit of $km \cdot mol^{-1}$.

| | Anharm. | Anharm. Int. | Obs. | Obs. Int. |
|---|---|---|---|---|
| $\nu_{17} + \nu_{21}$ | 1646 | 0.82 | 1645 (-1) | 0.06 |
| $\nu_{16} + \nu_{22}$ | 1654 | 0.28 | 1655 (1) | 0.03 |
| $\nu_{16} + \nu_{24}$ | 1709 | 0.85 | 1712 | 0.15 |
| $\nu_{19} + \nu_{22}$ | 1720 | 1.38 | | |
| $\nu_{17} + \nu_{23}$ | 1731 | 0.82 | 1729 (-2) | 0.11 |
| $\nu_{19} + \nu_{23}$ | 1759 | 0.43 | 1759 (0) | 0.05 |
| $\nu_{19} + \nu_{24}$ | 1774 | 0.20 | | |
| $\nu_{19} + \nu_{24}$ | 1786 | 0.40 | 1776 | 0.10 |
| $\nu_{21} + \nu_{22}$ | 1786 | 0.26 | | |
| $\nu_{21} + \nu_{23}$ | 1823 | 1.50 | 1821 | 0.18 |
| $2\nu_{22}$ | 1831 | 1.69 | | |
| $\nu_{21} + \nu_{25}$ | 1851 | 1.51 | 1855 (4) | 0.04 |
| $\nu_{22} + \nu_{23}$ | 1873 | 0.60 | 1870 (-3) | 0.05 |
| $\nu_{22} + \nu_{24}$ | 1888 | 1.69 | 1885 (-3) | 0.13 |
| $2\nu_{23}$ | 1912 | 0.31 | 1906 (-6) | 0.05 |
| $\nu_{23} + \nu_{24}$ | 1929 | 1.05 | | |
| $\nu_{23} + \nu_{25}$ | 1937 | 1.38 | 1933 | 0.19 |
| $2\nu_{24}$ | 1945 | 1.53 | | |
| $\nu_{24} + \nu_{25}$ | 1956 | 0.55 | 1954 | 0.12 |
| $2\nu_{25}$ | 1966 | 1.51 | | |

**Table A.3.** Experimental and B3LYP/N07D calculated vibrational frequencies and the vibrational assignments for 2-CNN. The numbers in parentheses indicate the difference between the experimental and harmonic frequencies. Frequencies labeled as Harm. and Obs. are in unit of $cm^{-1}$, while the harmonic intensities (Harm. Int.) are in unit of $km \cdot mol^{-1}$.

| | Harm. | Harm. Int. | Obs. | Obs. Int. | Assignment | Cal.[a] | FTIR [a] | Assignment [a] |
|---|---|---|---|---|---|---|---|---|
| $\nu_5$ | 349 | 2.40 | 351 (2) | 0.42 | α(CC) | 357 | 350 | α(CC) |
| $\nu_6$ | 382 | 0.01 | 377 (-5) | 0.18 | τ(CC) | 385 | 380 | τ(CC) |
| $\nu_7$ | 389 | 0.05 | 389 (0) | 0.15 | α(CC) | 393 | 393 | α(CC) |
| $\nu_8$ | 438 | 0.07 | — | — | τ(CC) | 445 | 438 | τ(CC) |
| $\nu_9$ | 473 | 20.05 | 470 (-3) | 0.97 | τ(CC) | 478 | 483 | τ(CC) |
| $\nu_{10}$ | 506 | 0.21 | 511 (5) | 0.15 | α(CC) | 514 | 509 | α(CC) |
| $\nu_{11}$ | 548 | 1.18 | 543 (-5) | 0.32 | τ(CC) | 558 | 548 | τ(CC) |
| $\nu_{12}$ | 560 | 1.23 | 560 (0) | 0.31 | α(CC) | 578 | 562 | α(CC) |
| $\nu_{13}$ | 617 | 2.64 | 616 (-1) | 0.37 | α(CC) | 629 | 617 | α(CC) |
| $\nu_{14}$ | 640 | 1.57 | 638 (-2) | 0.30 | τ(CC) | 653 | 643 | τ(CC) |
| $\nu_{15}$ | 671 | 0.01 | — | — | α(CC), R(CC) | 680 | 663 | — |
| $\nu_{16}$ | 740 | 33.60 | 740 (0) | 1.24 | ε(CH) | 751 | 755 | ε(CH) |
| $\nu_{17}$ | 752 | 0.48 | — | — | ε(CH), τ(CC) | 762 | 764 | α(CC), R(CC) |
| $\nu_{18}$ | 765 | 0.38 | 761 (-4) | 0.36 | β(CH), α(CC), R(CC) | 774 | 782 | ε(CH), α(CC) |
| $\nu_{19}$ | 812 | 37.18 | 810 (-2) | 1.14 | ε(CH), τ(CC) | 821 | 826 | β(CH), τ(CC) |
| $\nu_{20}$ | 855 | 16.30 | 854 (-1) | 0.95 | ε(CH), τ(CC) | 867 | 871 | ε(CH), τ(CC) |
| $\nu_{21}$ | 876 | 0.92 | — | — | α(CC), R(CC) | 886 | 884 | α(CC) |
| $\nu_{22}$ | 895 | 13.73 | 898 (3) | 0.79 | ε(CH) | 908 | 905 | ε(CH) |
| $\nu_{23}$ | 946 | 2.12 | 954 | 0.63 | ε(CH) | 958 | 960 | ε(CH) |
| $\nu_{24}$ | 950 | 3.47 | | | α(CC), R(CC) | 961 | 960 | α(CC) |
| $\nu_{25}$ | 960 | 1.94 | 960 (0) | 0.54 | ε(CH) | 972 | 967 | ε(CH) |
| $\nu_{26}$ | 974 | 0.01 | — | — | ε(CH) | 987 | 983 | ε(CH) |
| $\nu_{27}$ | 1021 | 2.68 | 1017 (-4) | 0.32 | β(CH), R(CC) | 1026 | 1018 | β(CH), R(CC) |
| $\nu_{28}$ | 1121 | 3.85 | 1124 (3) | 0.33 | β(CH), α(CC), R(CC) | 1132 | 1127 | β(CH), α(CC), R(CC) |
| $\nu_{29}$ | 1149 | 0.22 | — | — | β(CH), R(CC) | 1175 | 1147 | β(CH), R(CC) |
| $\nu_{30}$ | 1153 | 1.42 | 1159 | 0.47 | β(CH), R(CC) | 1166 | 1155 | β(CH), R(CC) |
| $\nu_{31}$ | 1160 | 4.35 | | | β(CH), R(CC) | 1171 | 1163 | R(CC) |
| $\nu_{32}$ | 1214 | 1.41 | 1209 (-5) | 0.25 | β(CH), R(CC) | 1203 | 1205 | β(CH), R(CC) |
| $\nu_{33}$ | 1249 | 0.95 | 1256 (7) | 0.26 | β(CH), α(CC), R(CC) | 1257 | 1248 | β(CH), α(CC) |
| $\nu_{34}$ | 1262 | 7.06 | 1268 (6) | 0.42 | β(CH), R(CC) | 1275 | 1273 | β(CH), α(CC), R(CC) |
| $\nu_{35}$ | 1348 | 0.79 | 1351 (3) | 0.22 | β(CH), R(CC) | 1353 | 1352 | β(CH), α(CC), R(CC) |
| $\nu_{36}$ | 1375 | 5.39 | 1363 (-12) | 0.33 | β(CH), R(CC) | 1364 | 1363 | α(CC), R(CC) |
| $\nu_{37}$ | 1385 | 2.67 | 1385 (0) | 0.28 | β(CH), R(CC) | 1373 | 1380 | β(CH) |
| $\nu_{38}$ | 1435 | 0.58 | 1435 (0) | 0.14 | β(CH), α(CC), R(CC) | 1444 | 1435 | β(CH), α(CC), R(CC) |
| $\nu_{39}$ | 1467 | 0.53 | 1467 (0) | 0.11 | β(CH), α(CC), R(CC) | 1471 | 1468 | β(CH), α(CC), R(CC) |
| $\nu_{40}$ | 1508 | 5.84 | 1504 (-4) | 0.41 | β(CH), α(CC), R(CC) | 1500 | 1501 | β(CH), α(CC), R(CC) |

**Table A.3.** continued.

| | Harm. | Harm. Int. | Obs. | Obs. Int. | Assignment | Cal.[a] | FTIR [a] | Assignment [a] |
|---|---|---|---|---|---|---|---|---|
| $\nu_{41}$ | 1575 | 0.39 | 1576 (1) | 0.17 | β(CH), α(CC), R(CC) | 1557 | 1568 | β(CH), α(CC), R(CC) |
| $\nu_{42}$ | 1607 | 2.94 | 1598 (-9) | 0.29 | β(CH), R(CC) | 1606 | 1594 | α(CC), R(CC) |
| $\nu_{43}$ | 1636 | 3.53 | 1633 (-3) | 0.17 | β(CH), R(CC) | 1624 | 1626 | α(CC), R(CC) |

Notes: R(CC): CC stretching; β(CH) and α(CC): CH and CC in-plane bending; ε(CH) and τ(CC): CH and CC out-of-plane bending; [a]These data were taken from Rawat et al. (2025)

**Table A.4.** Experimental and B3LYP/N07D calculated vibrational frequencies and the vibrational assignments (overtones and combination bands) for 2-CNN. The numbers in parentheses indicate the difference between the experimental and anharmonic frequencies. Frequencies labeled as Anharm. and Obs. are in unit of $cm^{-1}$, while the anharmonic intensities (Anharm. Int.) are in unit of $km \cdot mol^{-1}$.

| | Anharm. | Anharm. Int. | Obs. | Obs. Int. |
|---|---|---|---|---|
| $\nu_{13} + \nu_{27}$ | 1658 | 1.16 | 1655 (-3) | 0.11 |
| $\nu_{19} + \nu_{20}$ | 1675 | 0.93 | 1679 (4) | 0.05 |
| $\nu_{16} + \nu_{23}$ | 1689 | 0.57 | 1691 (2) | 0.18 |
| $\nu_{16} + \nu_{25}$ | 1703 | 0.58 | | |
| $\nu_{19} + \nu_{22}$ | 1715 | 0.15 | 1704 | 0.15 |
| $\nu_{17} + \nu_{23}$ | 1715 | 0.22 | | |
| $\nu_{17} + \nu_{25}$ | 1728 | 0.24 | 1728 (0) | 0.14 |
| $\nu_{17} + \nu_{26}$ | 1744 | 0.22 | 1741 (-3) | 0.14 |
| $\nu_{20} + \nu_{22}$ | 1759 | 0.33 | | |
| $\nu_{19} + \nu_{23}$ | 1761 | 1.44 | 1755 | 0.26 |
| $\nu_{19} + \nu_{25}$ | 1772 | 0.56 | | |
| $\nu_{19} + \nu_{26}$ | 1791 | 0.54 | 1787 | 0.20 |
| $2\nu_{22}$ | 1794 | 1.41 | | |
| $\nu_{20} + \nu_{23}$ | 1806 | 1.04 | 1807 (1) | 0.18 |
| $\nu_{20} + \nu_{25}$ | 1820 | 0.75 | 1822 (2) | 0.12 |
| $\nu_{20} + \nu_{26}$ | 1835 | 1.20 | 1832 (-3) | 0.12 |
| $\nu_{22} + \nu_{23}$ | 1847 | 0.24 | 1847 (0) | 0.09 |
| $\nu_{22} + \nu_{25}$ | 1860 | 0.48 | 1869 | 0.05 |
| $\nu_{22} + \nu_{26}$ | 1875 | 0.20 | | |
| $\nu_{23} + \nu_{25}$ | 1907 | 1.88 | 1902 (-5) | 0.21 |
| $2\nu_{25}$ | 1920 | 0.53 | 1922 | 0.18 |
| $\nu_{23} + \nu_{26}$ | 1923 | 1.09 | | |
| $\nu_{25} + \nu_{26}$ | 1937 | 0.92 | 1937 (0) | 0.11 |
| $2\nu_{26}$ | 1952 | 1.91 | 1947 (-5) | 0.20 |

## Appendix B: Simulated IR emission spectra of 1-CNN.

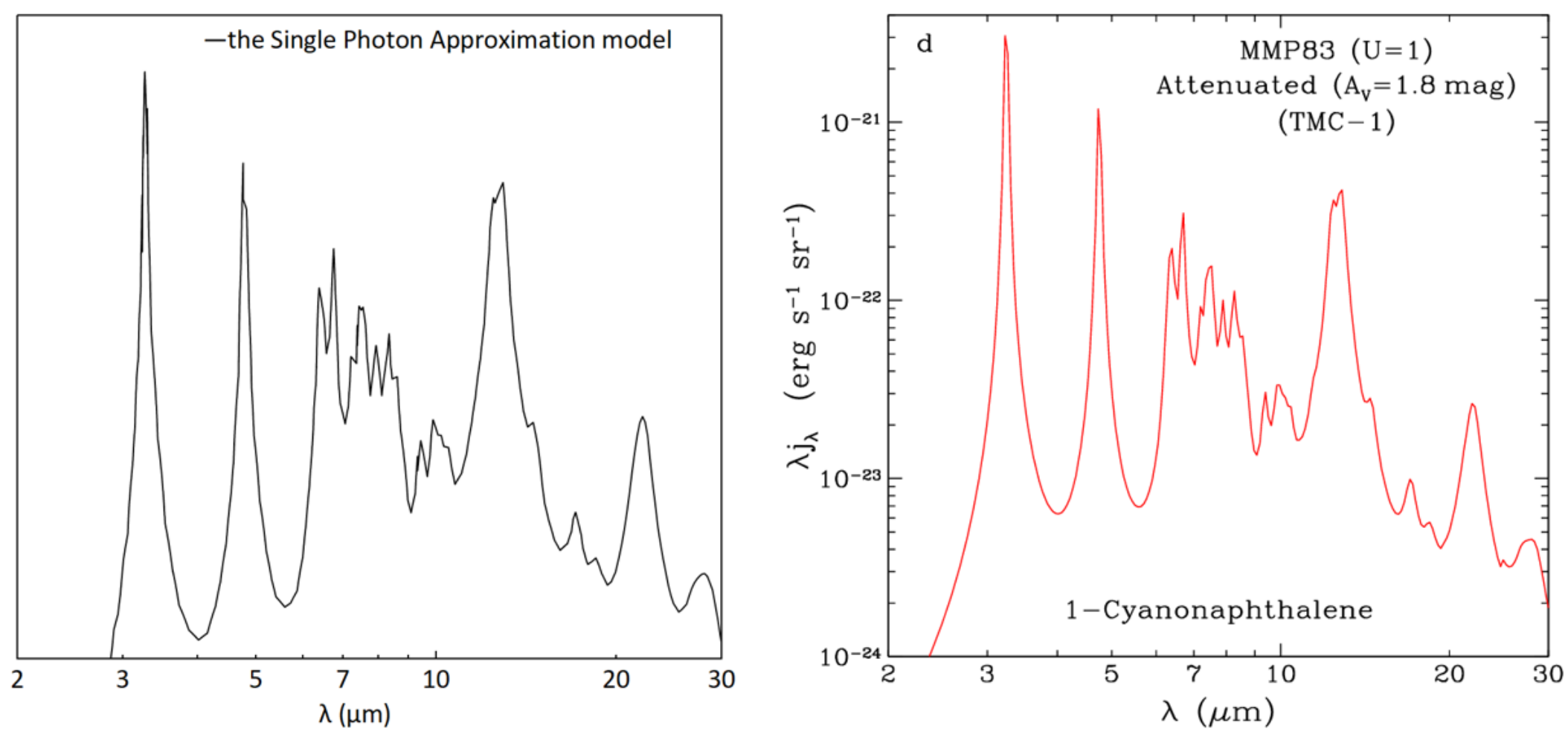

**Fig. B.1.** Comparison between the reproduced and published infrared emission spectra of 1-CNN. The reproduced spectrum (black) shown in the left panel was obtained using the Single Photon Approximation model and the absorption cross sections reported in the Li et al. (2024). The published spectrum (red) from Li et al. (2024) is shown in the right panel for comparison. The agreement between the two spectra validates the implementation of the emission model used in this work.